\documentstyle[12pt]{article}
\begin{document}

\vspace*{2cm}

\begin{center}
{\large\bf INVESTIGATION OF SELF-PRESERVATION THEORY IN 
TWO DIMENSIONAL TURBULENT MIXING LAYERS}
\end{center}

\vspace*{1.5cm}
\begin{center}
A. Shabani\footnote{arioshabani@yahoo.com, s7426959@cic.aku.ac.ir}
 and  H. Basirat Tabrizi\footnote{hbasirat@cic.aku.ac.ir} 
\end{center}

\vspace*{0.5cm}
\begin{center}
{\it Mechanical Engineering Department, \\
Amirkabir University of Technology, \\
Hafez St., Tehran, Iran}
\end{center}

\underline{\bf Abstract}

\noindent
The behavior of a two dimensional, steady turbulent mixing layer was 
investigated. Besides the usual 
velocity components, all the contributing components of the Reynolds Stresses are also calculated and 
presented. The results indicated that in the two dimensional steady turbulent mixing layers the theory of 
self-preservation is not valid for all the flow domain, and that the flow is turbulent regime is divided into 
two areas. Through calculation of Reynolds Stress components at the point of initiation of the flow, it was 
shown that the turbulent flow in the area containing the point of singularity of the flow is not, as previously 
believed, self-preserved. 

\noindent
\vspace*{1cm}

\section{Introduction} 

Turbulence is the most frequently occurring mode of fluid motion. It is characterized by 
high fluid particle mixing, and high energy dissipation [1-3]. The exact time-dependent 
nature of events leading to the onset and development of turbulence is not fully known; 
however, by the aid of time averaging techniques we can obtain meaningful and 
practically useful statistical definitions [3]. Nearly all the problems that are studied in the 
field of Turbulent Fluid Dynamics are numerically extensive and expensive [3-5], and 
there is no single concept to model different classes of turbulent flow regimes [3-5]. In 
general a turbulent flow is characterized by random variation of flow variables, high 
levels of fluid particle mixing, and high energy dissipation [2, 3]. 

Turbulent flow regimes may be divided into two main categories: Free Turbulence, 
where no physical boundary limits the development of flow: Bounded Turbulence, where 
flow is restricted by physical boundaries [2]. Experimentally, Free turbulent flows are 
easy to generate, and that is primarily why they have been so extensively studied. From a 
theoretical stand point, their formulation and simulation is also much simpler, since for 
normal incompressible flows, no pressure gradient term has relevance [3]. In studying 
free turbulent flow fields it is desirable for practical reasons to confine oneself, at first, to 
rather simple but important elementary flow patterns [1, 2]. These elementary flow 
patterns may be classified into two main groups: (1) free jets, and (2) wake flows behind 
obstacles. The flow pattern under investigation here is a plane, two dimensional mixing 
layer [3]. In this flow pattern it is possible to distinguish one main flow direction with 
velocity substantially greater than in any other direction [1, 2]. 
     
Through out the past few decades, turbulent plane mixing layer has been the subject of 
numerous experimental and theoretical research, and it is widely used to investigate 
various aspects of phenomena related to free turbulence, including process and nature of 
mixing [3], structure of turbulent components and their self-preservation [3-6], stability 
of the flow [4], transfer of momentum and energy [5], intensity, and rates of temporal and 
spatial expansions of free turbulent flows [6]. 

The spatial and temporal development of the turbulence is closely correlated to the 
development of Reynolds Stress Components. The existing literature indicates that for 
free turbulent flows, the entire flow domain was considered to be self-preserved [7]. 
However, in this investigation it is shown that the self-preservation theory is not 
applicable to all the flow field, and that at areas very close to the point of initiation of 
turbulent flow regime the said theory is not applicable, i.e. the flow is not self-preserved. 

Following the introduction the fluid dynamics equations are indicated and the 
formulation is completed. In section three, the simulation results and discussions are 
presented. And in the final section discussion and concluding remarks are presented.

\section{Turbulent plane mixing layer's governing equations}

For a 2-dimensional incompressible turbulent plane mixing layer, the governing 
equations are Reynolds equations and the continuity equation. They may be written in the 
following form:
\begin{equation}                                                                                                                               
\frac{\partial \bar u_1}{\partial t}+
\bar u_1 \frac{\partial \bar u_1}{\partial x}+
\bar u_2 \frac{\partial \bar u_1}{\partial y} =\nu
\frac{\partial^2 \bar u_1}{\partial y^2}-
\frac{\partial^2 \bar u_1'^{2}}{\partial x}-
\frac{\partial \bar {u_1'u'_2}}{\partial y}\;,
\end{equation}                                                                                                                               
\begin{equation}                                                                                                                               
\frac{\partial \bar u_2}{\partial t}+
\bar u_1 \frac{\partial \bar u_1}{\partial x}+
\bar u_2 \frac{\partial \bar u_2}{\partial y} =\nu
\frac{\partial^2 \bar u_2}{\partial y^2}-
\frac{\partial^2 \bar u_2'^{2}}{\partial y}-
\frac{\partial \bar {u_1'u'_2}}{\partial x} \;,
\end{equation}                                                                                                                               
\begin{equation}                                                                                                                               
\frac{\partial\bar u_1}{\partial x}+ 
\frac{\partial\bar u_2}{\partial y}=0\;,
\end{equation}                                                                                                                               
where bar represents time-averaged quantities, and $u_1$, $u_2$ are the velocity components in 
the $x$ and $y$ directions respectively; also $u_1'$, and $u_2'$ are the fluctuating components of the 
$u_1$ and $u_2$ velocity components respectively. 

To close the above set of equations the Reynolds Stress Equation Model ("RSM") is used 
[4]. The exact equation for the transport of Reynolds Stresses, $R_{ij}$, takes the following 
form:
\begin{equation}                                                                                                                               
\frac{\partial R_{ij}}{\partial t}+ div(R_{ij}V) = P_{ij} + D_{ij} - \varepsilon_{ij} +\Pi_{ij} +\Omega_{ij}\;,
\end{equation}                                                     
where $P_{ij}$, $D_{ij}$, $\varepsilon_{ij}$, $\Pi_{ij}$, and $\Omega_{ij}$ represent rate of production of 
Rij, transport of Rij by 
diffusion, rate of dissipation of $R_{ij}$, transport of $R_{ij}$ due to turbulent pressure-strain 
interactions, and transport of $R_{ij}$ due to rotation respectively. These terms are represented 
by the following equations:
\begin{equation}     
P_{ij}=-(R_{im} \frac{\partial \bar V_j}{\partial x_m}+  
R_{jm} \frac{\partial \bar V_i}{\partial x_m})\;,
\end{equation}     
where $V$  bar indicates the velocity vector, and having
\begin{equation}     
D_{ij}= \frac{\partial}{\partial x_m} (\frac{\nu_t}{\sigma_k} \frac {\partial R_{ij}}{\partial x_m})\;,
\end{equation}     
also,                                                                                                    
\begin{equation}   
\nu_t = (k^2 / \varepsilon ) C_\mu \;,
\end{equation}   
in which $C_\mu = 0.09$, and $\sigma_k = 1.00$.

Assuming isotropy of the small dissipative eddies, and assuming that it affects the normal 
Reynolds stresses $(i = j)$ only, and in equal measures we have [4]:
\begin{equation}   
\varepsilon_{ij} = (2 / 3) \varepsilon \delta_{ij}\;,
\end{equation}                              
where the Kronecker delta, $\delta_{ij}$ is given by $\delta_{ij} =1$ if $i = j$ and 
$\delta_{ij} = 0$ if $i \neq j$.

The transport of Reynolds stresses due to pressure-strain interactions is calculated by the 
following expression:
\begin{equation}   
\Pi_{ij} = - C_1 (\frac{\varepsilon}{k})(R_{ij} -(2 / 3) k \delta_{ij})-
C_2 (P_{ij} -(2 / 3) P \delta_{ij})\;,
\end{equation} 
in which $C_1 = 1.80$ and $C_2 = 0.60$.
The rotational term is given by:
\begin{equation}   
\Omega_{ij} = - 2 \omega_k (R_{jm} e_{ikm} + R_{im} e_{jkm})\;.
\end{equation}                                                                                                                                         
Here $\omega_k$ is the rotation vector and $e_{ijk}$ is the alternating symbol; $e_{ijk} = 1$ if $i$, $j$ and $k$ are 
different and in cyclic order, $e_{ijk} = -1$ if $i$, $j$ and $k$ are different and in anti-cyclic order and 
$e_{ijk} = 0$ if any two indices are the same. 

Turbulent kinetic energy $k$ is needed in the above formula and can be found by adding the 
three normal stresses together:
\begin{equation}   
k = 0.5 (R_{11} + R_{22} + R_{33})
\end{equation}
and the equation for the transport of scalar dissipation rate $\varepsilon$ is:
\begin{equation}
\frac{\partial (\rho\varepsilon )}{\partial t} +div (\rho\varepsilon V)=
div (\frac{\mu_t}{\sigma_\varepsilon } grad \varepsilon) +2 (\varepsilon / k) C_{1\varepsilon }
\mu_t E_{ij}\cdot E_{ij} -\rho (\varepsilon^2 / k) C_{2\varepsilon} \;,
\end{equation}
where $C_{1\varepsilon} = 1.44$ and $C_{2\varepsilon} = 1.92$, and $E_{ij}$ is 
the mean rate of deformation of a fluid 
element and $E_{ij}\cdot E_{ij}$ is their scalar product [4].
Note that for the free stream velocities $R_{ij} = 0$ and $\varepsilon = 0$.

Here the values of the above-defined relationships are presented. 
For $R_{12}$ we have:
\begin{equation}
P_{12} =- (R_{11}\frac{\partial \bar u_2}{\partial x_1}+ 
R_{12}\frac{\partial \bar u_1}{\partial x_1}+
R_{12}\frac{\partial \bar u_2}{\partial x_2}+
R_{22}\frac{\partial \bar u_1}{\partial x_2})
\end{equation}
and
\begin{equation}
D_{12} =\frac{\partial}{\partial x_1} (\frac{\nu_t}{\sigma_k} \frac{\partial 
R_{12}}{\partial x_1}) + 
\frac{\partial}{\partial x_2} (\frac{\nu_t}{\sigma_k} \frac{\partial 
R_{12}}{\partial x_2}) \;.
\end{equation}
Therefore,
\begin{equation}
D_{12} =\frac{\partial \nu_t}{\partial x_1}\frac{\partial R_{12}}{\partial x_1}
+\frac{\partial \nu_t}{\partial x_{21}}\frac{\partial R_{12}}{\partial x_2}+
\nu_t\frac{\partial^2 R_{12}}{\partial x^2_1}+
\nu_t\frac{\partial^2 R_{12}}{\partial x^2_2} \;,
\end{equation}                                                                                                                                        
where,
\begin{equation}
\frac{\partial \nu_t}{\partial x_1} =(0.09/\varepsilon )
[\frac{\partial (R_{11}+R_{22})^2}{\partial x_1}-
\frac{(R_{11}+R_{22})^2}{\varepsilon}\frac{\partial\varepsilon}{\partial x_1} ]
\end{equation}
and
\begin{equation}
\frac{\partial \nu_t}{\partial x_2} =(0.09/\varepsilon )
[\frac{\partial (R_{11}+R_{22})^2}{\partial x_2}-
\frac{(R_{11}+R_{22})^2}{\varepsilon}\frac{\partial\varepsilon}{\partial x_2} ]
\end{equation}
also,
\begin{eqnarray}
\Pi_{12} & = &(-1.8 \varepsilon /k) R_{12} -0.6 P_{12} \\
\Omega_{12} &= & 0 \\
\varepsilon_{12} &= &0 \;.
\end{eqnarray}                                                                                                                                         
For $R_{11}$ components we have:
\begin{eqnarray}
P_{11} &=& -2 (R_{11} \frac{\partial \bar u_1}{\partial x_1} +
R_{12} \frac{\partial \bar u_1}{\partial x_2})\;, \\
D_{11}&=& \frac{\partial}{\partial x_1} 
(\frac{\nu_t}{1.0}\frac{\partial R_{11}}{\partial x_1}) +
\frac{\partial}{\partial x_2} 
(\frac{\nu_t}{1.0}\frac{\partial R_{11}}{\partial x_2})\;, \\
\varepsilon_{11} &=& (2/3) \varepsilon \;, \\
\Pi_{11} &=& (-1.8 \varepsilon  /k ) (R_{11}- (2/3)k)-0.6 P_{11}\\
\Omega_{11}&=& 0 \;.
\end{eqnarray}                                                                                                                                        
And for $R_{22}$ components we have:
\begin{eqnarray}
P_{22} &=& -2 (R_{22} \frac{\partial \bar u_2}{\partial x_1} +
R_{22} \frac{\partial \bar u_2}{\partial x_2})\;, \\
D_{22}&=& \frac{\partial}{\partial x_1} 
(\frac{\nu_t}{1.0}\frac{\partial R_{22}}{\partial x_1}) +
\frac{\partial}{\partial x_2} 
(\frac{\nu_t}{1.0}\frac{\partial R_{22}}{\partial x_2})\;, \\
\varepsilon_{22} &=& (2/3) \varepsilon \;,\\
\Pi_{22} &=& (-1.8 \varepsilon  /k ) (R_{22}- (2/3)k)-0.6 P_{22}\;,\\
\Omega_{22}& =& 0 \;.
\end{eqnarray}                                                                                                                                        
For  $\varepsilon$ we have the following expressions:
\begin{eqnarray}
\frac{\partial \varepsilon}{\partial t}+ \bar u_1 \frac{\partial \eta}{\partial x_1}+
\bar u_2 \frac{\partial \varepsilon}{\partial x_2}= &&
\frac{\partial }{\partial x_1}(\frac{\nu_t}{1.3}\frac{\partial \varepsilon}{\partial x_1})+ 
\frac{\partial }{\partial x_2}(\frac{\nu_t}{1.3}\frac{\partial \varepsilon}{\partial x_2})+ 
\nonumber \\
&+& (\frac{2.88 \nu_t \varepsilon}{k})[E_{11}^2+2E_{12}^2+E_{22}^2]- 
\frac{1.92 \varepsilon^2}{k} \;,
\end{eqnarray}
where
\begin{eqnarray}
E_{11} &=& \frac{\partial \bar u_1 }{\partial x_1} \;, \\
E_{22} &=& \frac{\partial \bar u_2 }{\partial x_2} \;,\\
E_{12} &=& 0.5(\frac{\partial \bar u_1 }{\partial x_2} + 
\frac{\partial \bar u_2 }{\partial x_1}) \;,
\end{eqnarray}                                                                                                                                        
and the kinetic energy of turbulence is
\begin{equation}
k=0.5(R_{11}+R_{22}) \;.
\end{equation}

Therefore we have obtained six equations which must be solved simultaneously to 
provide the two unknown velocity components, three Reynolds stress components and 
the scalar turbulent dissipation rate.  

With the aid of standard Computational Fluid Dynamic (CFD) schemes and formulations, 
namely the Central Finite Difference Discretization, and using a forward time-marching 
scheme, the equations are numerically simulated and solved. The time marching is 
carried out until the steady state conditions were established. The $x$, $y$ and $t$ steps are 
determined by trial and error and are fixed and equal to $0.0001/40$ m, $0.00001/40$ m, 
and $10^{-7}$ seconds respectively.

\section{Simulation results and discussions}

The respective computer codes for the CFD formulation of the turbulent plane mixing 
layer are run until the steady state solution is obtained. Using the approximation, for free 
stream velocities of $U_{max} = 10.0 (m/s)$, $U_{min} = 5.0 (m/s)$, the velocity component 
distributions and the non-dimensional Reynolds stress components at various flow 
sections are obtained, and for sections located at about $x = 0.00001 (m)$, near the 
initiation point of the flow, where inherently a critical point of discontinuity exists, and 
about $x = 0.000036 (m)$, far down-stream from the point of initiation of the flow, are 
plotted and presented in the following figures. In the following figures $U_{ref} = U_{max} - U_{min}$.

Figure 1 indicates the experimental results for the spatial expansion of the turbulent 
mixing layer obtained by D. Oster and I. Wygnanski [7]. In the same figure, the 
simulation results obtained from the utilized CFD code are indicated.

In Figure 2, the Reynolds Stress distribution obtained experimentally [7], at $x = 300 mm$, 
$x = 500$ mm through 1700 mm, and by the utilized CFD formulation, at the area close to 
the point of initiation of the turbulent flow, is shown.  
 
Figures 3 and 4 indicate the variation of u velocity component against the non-
dimensional vertical distance at two intersections close to the point of initiation of the 
mixing layer, as calculated by the utilized CFD scheme.

Figures 5 and 6 show the distribution of non-dimensional $u'^2$ Reynolds Stress 
Components at cross sections close to the point of initiation of turbulent flow, calculated 
by the used CFD scheme.

From these figures, the reduction in the peak value of turbulent intensity in the flow 
direction indicates that the in the region under consideration the flow is not self-
preserving. 

Figures 7 and 8 present the variation of the non-dimensional $v'^2$ Reynolds Stress 
Components at cross sections close to the point of initiation of turbulent flow, determined 
by the used CFD code. 

Figures 7 and 8 also indicate that the turbulent flow at the desired area is not self-
preserving. In addition, comparing the order of magnitudes of $u'^2$ and $v'^2$ Reynolds Stress 
components in figures 8 through 10, although the relative contribution of $u'^2$ component 
to the generation of turbulence is more, its effect on further development of turbulence 
regime downstream of the point of initiation of the plane mixing layer is reduced. 

Figures 9 and 10 show the distribution of the time averaged, non-dimensional auto-
correlated components of the turbulent velocity fluctuations, very close to the point of 
emergence of the mixing layer, obtained by the CFD solution.          

From figures 9 and 10, comparing the maximum value of the turbulence intensity with 
the experimental values measured by Oster and Wygnanski [7], Yule, and Spencer, both 
presented at the same reference and indicated in Table 1, the acceptable performance of 
the CFD solution and our results is established.

\begin{center}
{\small {\bf Table 1.} The experimental and theoretical maximum value of
 intensity \\ for $U_{max}/U_{min}$ ratio of 0.5.}
\end{center}
\begin{center}
\begin{tabular}{|c|c|} \hline
Source & $\overline{ u'v' }/U_{ref}^2$ \\ \hline 
Spencer (1970) & 0.011   \\ \hline
Yule (1971) & 0.013   \\ \hline
Oster and Wygnanski (1982) & 0.013 \\  \hline 
Present result & 0.013 \\ \hline
\end{tabular}
\end{center}

In addition, Figures 5 through 10 indicate that turbulence Reynolds Stresses, which are 
themselves responsible for turbulence generation, have their peak value at the center of 
symmetry of the flow. This fact together with the order of magnitude of Reynolds Stress 
variation complies with the known behavior of such turbulent flow regimes [3-7]. 

Based upon their experimental set up, erected to study free and forced turbulent plane 
mixing layers, D. Oster and I. Wygnanski had deducted that for free stream velocity 
ratios of less than 0.6 the flow remains self-preserving [7]. However, their area of interest 
started 200 mm down stream of the point of initiation of the turbulent flow.

The results obtained by the utilized formulation indicated that for the area of interest in 
such flow patterns the auto-correlated Reynolds Stress components have closer spatial 
behavior, but the cross-correlation of turbulent velocity components has a different 
contribution to the generation of turbulence, as it shows a larger peak at 0.00001 m, and 
decays faster at 0.000036 m from the point of initiation of turbulent flow regime.

Referring to the our results, D. Oster and I. Wygnanski's deduction is invalid for the area 
of the flow close to the point of initiation of turbulence regime. In addition, figures 5 
through 10 indicate that at points further away from, but close to, the point of initiation of 
turbulence flow, the auto-correlated components of Reynolds Stresses play the dominant 
role in generating and maintaining the turbulence behavior. As a further expansion to 
such deduction it can be noted that at initial stages of such flow regimes, turbulence starts 
and expands more due to contribution of cross-correlated component of velocity 
fluctuations, and as the influenced region of turbulent flow increases, most of the 
generated momentum and energy of turbulent components needed to generate and 
maintain the turbulent flow pattern is supported through the influence of the auto-
correlated fluctuating velocity components. In other words, there are two regions of 
Reynolds Stress contribution to the generation and expansion of turbulence. In the first 
region, the turbulence is generated through a higher correlation of vertical fluctuating 
velocity components; such close correlation decays yielding a second mechanism of 
turbulence generation, forming a region at which turbulence is mainly maintained by 
higher temporal auto-correlation of fluctuating velocity components.

\section{Conclusion}

The particular area under consideration includes a point of discontinuity, critical point, 
where all the flow variables are constantly zero [3-7]. As the starting point of the 
turbulent flow regime, the consequent development of the turbulent mixing layer is 
dependent to events that take place in this very small region. Comparing the results with 
similar experimental research on free turbulent plane mixing layers, we deduct that the 
obtained results are acceptable [5, 7]. However, the complex nature of events in this 
region and their ultimate effect on the development of turbulent flow deserve much more 
theoretical and experimental investigation.

\newpage

\section*{References}

\begin{description}
\item[1.] Schlichting, H., 1968, Boundary layer theory, 
McGraw Hill Series.
\item[2.] Hinze, J. O., 1959, Turbulence, McGraw Hill Series.
\item[3.] Grinstein, F. F., Oran, E. S., Boris, J. P., 1986, 
Numerical simulations of mixing in 
planar shear flows, Journal of Fluid Mechanics, Vol. 165, 201-220.
\item[4.] Hartel, C., 1996, Turbulent flows: 
direct numerical simulation and large-eddy 
simulation, Handbook of Computational Fluid Mechanics, Academic Press Limited.
\item[5.] Versteeg, H. K., Malalasekera. W., 1995, 
An introduction to computational fluid 
dynamics, the finite volume method, Addison Wesley Longman Limited.
\item[6.] Fazle Hussain, A. K. M., 1986, 
Coherent Structures and Turbulence, Journal of 
Fluid Mechanics, Vol. 173, 303-356.
\item[7.] Oster, D., Wygnanski, I., 1982, 
Forced mixing layer between parallel streams, 
Journal of Fluid Mechanics, Vol. 123, 91-130.
\end{description}

 \newpage

\section*{Figure captions}

\begin{description}
\item[Fig. 1.]  Comparison of the spreading of 
unforced mixing layer obtained by the utilized 
CFD and experimental set up of D. Oster and I. Wygnanski. 
\item[Fig. 2.] Distribution of time averaged 
$\overline{|u'v'|}/U^2_{ref}$ 
 for $U_{max}/U_{min}$ of less than 0.6, provided 
by D. Oster and I. Wygnanski, and calculated by the utilized CFD scheme.
\item[Fig. 3.] Graph of Velocity Component $u$, as calculated by the CFD 
scheme for flow cross-section at $x =      0.00001$ m, 
for the free stream velocity ratio of 1/2.
\item[Fig. 4.] Graph of velocity component $u$, as calculated by the CFD 
scheme for flow cross section at $x =      0.000036$ m, 
for the free stream velocity ratio of 1/2.
\item[Fig. 5.] Graph of variation of average 
$\overline{ |u'u'|}/U_{ref}^2$, 
against vertical distance, as calculated by 
CFD scheme, for flow cross section at $x = 0.00001$ m, 
for free stream velocity ratio of 1/2.
\item[Fig. 6.] Graph of variation of average 
$\overline{ |u'u'|}/U_{ref}^2$, 
against vertical distance, as calculated by 
CFD scheme, for flow cross section at $x = 0.000036$ m, 
for free stream velocity ratio of 1/2.
\item[Fig. 7.] Graph of variation of average 
$\overline{ |v'v'|}/U_{ref}^2$, 
against vertical distance, as calculated by 
CFD scheme, for flow cross section at $x = 0.00001$ m, 
for free stream velocity ratio of 1/2.
\item[Fig. 8.] Graph of variation of average 
$\overline{ |v'v'|}/U_{ref}^2$, 
against vertical distance, as calculated by 
CFD scheme, for flow cross section at $x = 0.000036$ m, 
for free stream velocity ratio of 1/2.
\item[Fig. 9.] Graph of variation of average 
$\overline{ |u'v'|}/U_{ref}^2$, 
against vertical distance, as calculated by 
CFD scheme, for flow cross section at $x = 0.00001$ m, 
for free stream velocity ratio of 1/2.
\item[Fig. 10.] Graph of variation of average 
$\overline{ |u'v'|}/U_{ref}^2$, 
against vertical distance, as calculated by 
CFD scheme, for flow cross section at $x = 0.000036$ m, 
for free stream velocity ratio of 1/2.
\end{description}
\end{document}